From dfsfgg8@PS.UIB.ES Fri Jan  2 10:37:08 1998
X400-Received: by /PRMD=iris/ADMD=mensatex/C=es/; Relayed; Fri,  2 Jan 1998 10:25:16 UTC+0100
Date: Fri,  2 Jan 1998 10:25:16 UTC+0100
X400-Originator: dfsfgg8@PS.UIB.ES
X400-Recipients: non-disclosure:;
X400-Content-Type: P2-1984 (2)
X400-MTS-Identifier: [/PRMD=iris/ADMD=mensatex/C=es/;980102102516]
Content-Identifier: 3553
From: Francesca Garcias <dfsfgg8@PS.UIB.ES>
To: Martin Garcia <garcia@physik.fu-berlin.de>
Subject: PRBtex
MIME-Version: 1.0 (Generated by Ean X.400 to MIME gateway)
Content-Length: 27206

\documentstyle[aps,preprint]{revtex}
\begin{document}
\normalsize
\draft
\title{ Nonradiative Electronic Deexcitation Time Scales in Metal Clusters}
\author{M. E. Garcia$^*$, Ll. Serra, F. Garcias,} 
\address{\it Departament de F\'{\i}sica, Universitat de les Illes 
Balears, E-07071 Palma de Mallorca, Spain}
\author{and K. H. Bennemann}
\address{\it Institut f{\"u}r Theoretische Physik, Freie Universit{\"a}t 
Berlin, Arnimallee 14, 14195 Berlin, Germany }
\date{\today }
\maketitle
\widetext

\begin{abstract}
The life-times due to Auger-electron emission  for a hole on a deep electronic 
shell   of   neutral and charged                         
 sodium clusters are studied for different sizes. We consider spherical 
clusters and calculate the Auger-transition probabilities using the energy 
levels and wave functions calculated in the Local-Density-Approximation (LDA).
  We obtain that    
Auger emission processes are energetically not  allowed for neutral and 
positively charged sodium clusters.  
In general, the Auger probabilities in  small 
Na$_N^-$ clusters are remarkably  
different  from the  atomic ones and exhibit  
a rich size dependence. 
 The Auger decay times of  most of the cluster sizes 
studied  are orders of magnitude larger than in atoms and might  be  
 comparable with typical fragmentation  times.   
\end{abstract} 

\newpage 
\noindent
\section{Introduction}
The decay of an electronically excited cluster may take place by emission 
of electrons, atoms or photons.  It is generally accepted that the radiative 
cooling is the decay channel with the largest time scale\cite{recknagel}.
 Another deexcitation channel is the evaporation  of atoms, which   
 occurs within a 
 characteristic time which ranges 
between pico- and milliseconds, depending on the bonding character of the 
clusters and the excitation energy. 
 The fastest relaxation channel is the emission of photoelectrons. 
 In addition, and like in atoms,  secondary electrons can be emitted if the 
excitation energy is larger than the ionization potential of the cluster. 
 These electrons result from intraband Auger processes, i.e., Auger decay of 
 a valence hole and subsequent emission of a valence electron.  

In excited light atoms, the nonradiative decay through emission 
of Auger electrons has a much larger probability than the radiative decay and 
 dominates the relaxation process\cite{bambynek}. 
 In clusters there are so far neither experimental nor theoretical studies 
of the intraband Auger probabilities. Recently, two important  experimental 
 studies on  the 
deexcitation channels of small clusters after optical excitation have been 
performed, from which the contribution of Auger-processes can be inferred.  
 Gantef\"or et al \cite{gantefor} analyzed the kinetic energy distribution 
of electrons emitted from optically excited clusters. They found that the 
photoelectron spectra show contributions from three different processes: 
direct emission, thermionic emission, and ``inelastic scattered electrons''. 
 The latter appear within a definite and narrow range of kinetic energies. 
 One could in principle interpret the inelastic scattered electrons as 
coming from intraband Auger-processes. 
 Reiners and Haberland\cite{haberland} studied the competition between 
 electron and atom emission after photoabsorption as a function of the 
photon energy. They observed that atom emission occurs within an energy
 range which is smaller than the cluster bandwidth. This means that 
excitations consisting of  a hole  in the bottom of the valence band 
(i.e., in the deepest electronic shells) do not lead to atom emission, 
despite that their energy is larger than the binding energy. The authors
 conjectured that this is due to intraband Auger-processes which lead to
 a very short life-time of such holes. 

 In this paper we present the first calculations of the intraband 
Auger time scales in metal clusters and show that Auger  emission
 probabilities strongly depend on size and are, in general, orders of 
magnitude smaller than atomic Auger probabilities. 

\section{Theory}
We consider only closed-shell Na$_N$, Na$_N^+$ and Na$_N^-$ clusters and make
 use of their spherical symmetry. For the description of the electronic 
structure of the clusters we use the jellium model\cite{walter}.   
Through the
 Auger process, a vacancy in a state $|n'' \ell'' \rangle$, with 
$\varepsilon_{n''\ell''} \le \varepsilon_F$ ($\varepsilon_F$ being the Fermi 
level), is filled by an electron coming from a higher bound 
 level $n\ell$. The energy 
released by this transition is transferred to a second electron, initially 
in a bound state  $n'\ell'$,  
 which is ejected (i.e., excited into a continuum state $k$). 
 Thus, the initial state consists of one hole in a bound state, and 
 the final state consists of two holes in bound states and one 
electron in a continuum state. One can, however, consider the Auger 
process as a two hole $\rightarrow$ two hole transition, where the initial
 state is given by $| \Psi_i  \rangle = c_{n'' \ell'' \sigma} \; c_k \; 
|\Psi_k \rangle$,  with $|\Psi_k \rangle = c^+_k  |\Psi_0 \rangle$ and  
$\varepsilon_{k} > 0$, and the final state by $| \Psi_f  \rangle = 
 c_{n \ell\sigma}\; 
c_{n' \ell' \sigma'}\;|\Psi_k 
\rangle$, with $\varepsilon_{n' \ell'},\; \varepsilon_{n \ell}  \le 
\varepsilon_F$. $k$ denotes the continuum state of the Auger electron and
 $n'' \ell'', n\ell$ stand for the states corresponding to the two final 
holes. $ |\Psi_0 \rangle$ refers to the electronic ground state of the 
cluster. The Auger transition probability can be calculated using Fermi's 
golden rule, and is  given by 
\begin{equation}
\label{golden}
 w_{fi} = \frac{2 \pi}{\hbar}  \; \left| \langle \Psi_f \mid 
{\hat V} \mid 
 \Psi_i \rangle \right|^2 \; \rho(E_f),
\end{equation}
 where $\rho(E_f)$ corresponds to the density of final states and ${\hat V}$
 is the operator describing the Coulomb interactions, which is written as
\begin{equation}
\label{coul}
{\hat V} = \frac{1}{2} 
   \; \, \sum_{1234\atop\sigma_1\sigma_2  }  
 V_{1234} \; c^{+}_{1 \sigma_1}\, c^{+}_{2 \sigma_2} \,  
c^{}_{3 \sigma_2} \,
  c^{}_{4 \sigma_1},      
\end{equation}
where the sum is over the cluster energy levels. 
In Eq.~(\ref{coul}) the quantities $V_{1234}$ are  the Coulomb matrix elements 
\begin{equation}
\label{matel}
V_{1234} = \int \int \psi^*_1({\vec r\,}) \psi^*_2(\vec{r}\, ') \; 
 \frac{e^2}{|{\vec r} -{\vec{r}\, '}|}  \;  
\psi_3(\vec{r}\, ') \psi_4({\vec r}\,) \; d{\vec r} \;d{\vec{r}\, '} , 
\end{equation}
expressed in the basis of eigenfunctions corresponding to the 
  bound and continuum levels of the cluster. 

In order to calculate the Auger-emission probability of Eq.~(\ref{golden}) we
have first to determine the wave functions of the three bound states and 
that of the continuum state which is involved in. This can be done by
performing  an  
 approximation  which is widely used 
in atomic physics\cite{bambynek}, and consists in taking the wave 
functions and energy
 levels of the original system, i.e., the cluster before the creation of the
initial vacancy.   
 Therefore we solve  
 the Kohn-Sham equations for the spherical metal cluster, given by 
\begin{equation}   
\label{ks}
\left[ - {1\over 2 r} { {\rm d}^2 \over {\rm d} r^2} r +
{ \ell(\ell+1) \over 2 r^2}  + V_{\em eff}(r) \right] R_{n\ell}(r) = 
\varepsilon_{n\ell} R_{n\ell}(r).  
\end{equation}
Here,  $V_{\em eff}(\vec r\,)$ is the effective one-electron potential
\begin{equation}
V_{\em eff}(\vec r\,) = v_j(\vec r\,) + \int {n(\vec{r}\, ') \over 
\left\vert \vec r - \vec {r}\, ' \right\vert}\, {\rm d}\vec {r}\, '  +
{\delta\over \delta n} E_{XC}[n(\vec r\,)] \; ,
\end{equation}
where  $v_j(\vec r\,)$ is the  electrostatic potential created by 
the jellium distribution of charge and $E_{XC}[n(\vec r\,)]$
the exchange and correlation term\cite{baladron}.
$n(\vec r\,)$ is the electronic density.  
 From Eqs.~(\ref{ks}) one obtains the bound states of the cluster. 
 The numerical algorithm to solve  Eqs.~(\ref{ks}) imposes the condition of 
regularity at the 
origin $R_{n\ell}\propto r^{\ell}$ and fixes the number of nodes 
$n$ of $R_{n\ell}$.
There is only one solution for a fixed $n$ and $\ell$ which 
vanishes exponentially 
at infinity and the algorithm iterates to find it, as well as 
its associated eigenvalue $\varepsilon_{n\ell}$. 
The number of occupied shells determines the number of possible 
Auger processes for a given hole in the shell $n'' \ell''$. 
 Since the total energy is conserved during the transition ($E_i = E_f$), 
the kinetic energy 
$\varepsilon_{k \ell_k}$ of the emitted Auger electron is given by 
$\varepsilon_{k \ell_k} = 
|\varepsilon_{n'' \ell''}| - |\varepsilon_{n' \ell'}| - 
|\varepsilon_{n \ell}|$.  
For a Auger emission to take place it must obviously hold that 
$\varepsilon_{k \ell_k}  > 0$. Since also the total angular momentum is 
conserved, the two-hole final 
state must have the same angular momentum as the two hole initial state 
$(L_i = L_f = L)$. 
 This requires for the angular momentum $\ell_k$ of the Auger electron 
the condition  
 $ |L-\ell''| \le \ell_k \le L+\ell''$, where $L$ must satisfy the inequality 
$|\ell -\ell'| \le L \le \ell + \ell'$. Similar constraints are fulfilled by
 the 
spin of the Auger electron.   
 For the calculation of the Auger continuum wave function 
 we also use  Eqs.~(\ref{ks}). However, in this case 
 the number of nodes is unknown and  the continuum energy
$\varepsilon_{k \ell_k}$ and the multipolarity $\ell_k$ are fixed. 
The same regularity condition is used to start the integration outwards, 
from the 
origin up to a fixed
large radius $R_0$. We normalize the outgoing Auger wave function 
$\sim \exp(ikr)/r$  within a sphere 
of radius $R_0$. Thus, the density of final states is given by
 $\rho(E_f) = R_0/(2\pi \hbar v)$, where  $v$ refers to 
the velocity of the Auger electron \cite{bambynek}. 

In order to determine the Auger probability we first separate 
the matrix elements (\ref{matel}) into radial and angular factors. 
This is achieved by  performing  the multipole 
expansion 
$ 1/r_{12} = \sum_{\mu \nu} \, r^{\nu}_< \; / \; r^{\nu}_>  \,  
C^*_{\mu \nu}(\theta_1, \phi_1) 
 \, C_{\mu \nu}(\theta_2,\phi_2 )$, with 
$ C_{\mu \nu} = \sqrt{\frac{4\pi}{2\nu + 1}} \; Y_{\mu \nu}(\theta, \phi) $, 
being  $Y_{\mu \nu}(\theta, \phi)$ spherical harmonics. 
 Evaluation of the angular factors depends on the choice of the 
angular-momentum coupling scheme. Since for the valence electrons of clusters 
 the spin-orbit coupling is negligible, the initial and final two-hole states 
of the cluster can be expressed in the 
 (LSJM) representation.  The total transition probability into all 
possible states of $L$ and $S$ is then given by\cite{bambynek}
\begin{equation}  
\label{prob}
w(n \ell,n' \ell',n'' \ell'') = 
\sum_{L,S} \frac{(2S+1)(2L+1)}
{2(2\ell''+1)}   \sum_{\ell_k} \left|
\frac{1}{2\hbar} \sum_{\nu} \left[ d_{\nu} D_{\nu} \; + \; (-1)^{L+S+\ell
+\ell'}\; \; e_{\nu} J_{\nu} \right] \right|^2 
  \rho(E_f) 
\end{equation}
where the functions $D_{\nu}$ and $J_{\nu}$ are the direct and 
exchange radial matrix elements, and the angular factors $d_{\nu}$ 
and $e_{\nu}$ are given by 
\begin{equation}
d_{\nu} = (-1)^{\ell+\ell'+L} \; \langle \ell'' || C^{\nu} || \ell \rangle \; 
\langle \ell_k || 
C^{\nu} || \ell' \rangle \; \left\{ \begin{array}{ccc} \ell'' & \ell_k & L \\
\ell' & \ell & \nu \end{array} \right\} 
\end{equation}
and 
\begin{equation}
e_{\nu} = (-1)^{\ell_k+\ell'+L} \;  \langle \ell'' || C^{\nu} || \ell' \rangle 
\; \langle \ell_k || C^{\nu} || \ell \rangle \; \left\{ \begin{array}{ccc} 
\ell'' & \ell_k & L \\
\ell & \ell' & \nu \end{array} \right\}.  
\end{equation}
Here, $ \langle \ell || C^{\nu} || \ell' \rangle $ is the reduced matrix 
element of the spherical harmonic, multiplied by 
$[4\pi/(2\nu + 1)]^{1/2}$\cite{bambynek}. 

\section{Results and Discussion}
We have calculated the Auger transition probabilities for spherical 
(closed-shell) 
Na$_N$, Na$_N^+$ and Na$_N^-$ clusters with $19 \le N \le 253$.  For 
each cluster we have determined first the effective potential 
$V_{\em eff}(\vec r\,)$, the bound 
Kohn-Sham states $(\{\varepsilon_{n\ell}\} < 0)$ and the corresponding radial 
eigenfunctions $\{R_{n\ell}(r)\}$. For a given vacancy $n'' \ell''$ we 
calculated the number of possible final states $\{n\ell, n'\ell'\}$. For 
each of 
the final states   we determined 
 the energy $\varepsilon_k$ of the 
emitted electron and the different possible values of 
its angular momentum $\ell_k$.  
 Then, for each $\ell_k$, we determined the corresponding outgoing radial 
wave function. 
This allowed us to calculate  
the Coulomb matrix elements and, using Eq.~(\ref{prob}), the  
 probability $w(n \ell,n' \ell',n'' \ell'')$. Finally,
 we have calculated the total probability for an initial vacancy $n'' \ell''$ 
as 
 $W(n'' \ell'') = \sum_{\ell \ell'} w(n \ell,n' \ell',n'' \ell'')$. The total 
probabilities $W(n'' \ell'')$ can,
 of course,  also be expressed as an energy width [$\Sigma_A(n'' \ell'') = 
\hbar/W(n'' \ell'')$] or as a life-time [$\tau_A(n'' \ell'') = 
W(n'' \ell'')^{-1}$]. 

 The first interesting result of our study 
is that, within the approximation used in Eq. (5) for the exchange and
correlation term (LDA), for neutral and positively 
charged spherical sodium clusters 
intraband Auger processes are energetically forbidden. There is no 
possible transition 
 $n''\ell'', n\ell , n'\ell'$ yielding $\varepsilon_k > 0$. This means that, 
according to our LDA calculations,   excited 
 spherical Na$_N^+$ and Na$_N$ clusters can only decay via fragmentation 
(evaporation) or photon emission.  

 Our results for negatively charged 
sodium clusters indicate that, 
 in contrast to what occurs for Na$_N^+$ and Na$_N$,   
 nonradiative electronic decay is possible. Due to the presence of the 
extra electron and the consequent extra Coulomb repulsion, the binding energy
 of the electrons in negatively charged clusters is smaller than in neutral or
 positively charged ones. As a consequence the whole band of bound states is 
shifted upwards and makes possible Auger transitions with $\varepsilon_k > 0$. 
 An example of such Auger transitions for negatively charged sodium clusters 
is illustrated schematically in Fig.~1.   Note that the Kohn-Sham effective 
potential  $V_{\em eff}(r)$ shows a barrier for negatively charged clusters. 
This might have consequences for the magnitude of the nonradiative emission 
probabilities. For instance, the wave function of an  
emitted Auger-electron with positive energy 
 but smaller than the energy barrier could have a  large weight 
inside the cluster due to trapping effects and influence the transition 
matrix elements used to calculate $W$. 
 For holes in the first two shells of Na$_N^-$ there are many possible 
transitions. For instance, there are 6 ways of filling a $1s$-vacancy in
Na$^-_{39}$ by emitting an electron. For Na$^-_{91}$ the number of such 
transitions is 11; for Na$^-_{137}$, 14, and for Na$^-_{253}$, 19. 
 
Our calculated  Auger widths of  Na$_N^-$ clusters are, 
for some cluster sizes, of the order of $10^{-1}$ eV, i.e., as large as 
for light atoms.  
In Fig.~2  the Auger life-time of a hole in the first $(1s)$ and second 
$(1p)$ shells of spherical 
Na$^-_N$ clusters is shown  as a function of the cluster size. 
  The size dependence of $\tau_A(1s)$ and $\tau_A(1p)$ is very rich and shows
 no monotonical behavior. Furthermore, the life-times oscillate over many 
orders of magnitude. For instance,  an initial $1s$-hole  
in excited Na$_{39}^-$ and Na$_{67}^-$ lives only few femtoseconds,  
almost as short as a $1s$-vacancy in light atoms with 
$Z < 10$\cite{bambynek}. For these 
clusters the Auger emission is faster than any other deexcitation mechanism.
 This means that one can separate the time scales for the electronic and 
atomic relaxation. Thus, if one is interested in  studying, for instance, 
 the atomic motion after optical excitation of Na$_{39}^-$ and Na$_{67}^-$, 
one can assume that the Auger process occurs immediately after the excitation 
and has no further influence on the fragmentation (evaporation) behavior. 
 On the other hand, $\tau_A(1s)$ for Na$_{57}^-$ and  $\tau_A(1p)$ for 
Na$_{39}^-$ and
 Na$_{57}^-$ are remarkable large, of the order of nanoseconds, i.e., 
larger that the life-times for any other deexcitation channel. 
 For this other extreme case one can again separate the time scales and 
assume that the hole lives infinitely long (compared with the atomic 
relaxation). 

 However, it is not always possible to perform this separation of 
time scales. 
 Fig.~2 also shows that for most cluster sizes $\tau_A$ lies between 
pico- and nanoseconds. This magnitudes are  comparable to the Auger 
life-times of positively charged ions approaching a metal surface at a 
distance of at least 2 \AA \cite{lorente}.  
This time scale is probably in the range in which excited clusters
 fragment\cite{brechignac}. Thus, results of Fig.~2 suggest that there 
might be competition between  Auger-emission and fragmentation channels for
a vacancy in a deep shell of a Na$_N^-$ cluster.
 
In Fig.~3 the Auger life-time for initial vacancies in the different 
electronic shells of Na$_{137}^-$ are shown.  The solid line shows results 
obtained from the calculations as described before, whereas the dashed curve
 shows results obtained including relaxation effects in the final state 
(shake-off). 
 The final state of the cluster corresponds  actually to a system  
with $N-1$ electrons. This should have an influence on the 
Kohn-Sham levels, reflecting the fact that the many electron system relaxes 
due to the excess nuclear positive charge.  
 For the cluster sizes considered, which have closed shells in the initial
 state, the final state with $N-1$ electrons is no longer a closed-shell 
system and   cannot be calculated using the spherical 
jellium model\cite{ekardt}. Thus, in order to take into account 
the relaxation effects we performed the following approach, commonly used 
in atomic physics\cite{bambynek}. We considered
 for both the initial and the final state a cluster with $N$ electrons, i.e.,
 a closed-shell configuration. However, for the final state we solved the
 Kohn-Sham problem for a positive jellium background with charge 
$Q = |e| (N+1)$.  In this way we simulated the excess charge. 
 The relaxation effects calculated within this approach 
do not change the qualitative trend of 
the results, neither for the size dependence nor for the shell dependence 
(for fixed size) of the Auger life-times, as seen in Fig.~3. 
 Regarding the shell-dependence of $\tau_A$ for the different cluster sizes 
studied, there is no clear dependence on the level of the initial vacancy. 
$\tau_A(1s)$ is in most cases the smallest life-time. For some clusters  
 $\tau_A(m)$ shows an alternation for increasing shell number $m=1,2,..$, 
like for Na$_{137}^-$. For other cluster sizes there is a monotonic increase 
of $\tau(m)$ with $m$.  
In general one would expect a monotonic increase of the life-time, since the 
 Coulomb matrix elements [Eq.~(\ref{matel})] should decrease with 
increasing kinetic energy of 
the Auger electron. However, due to the potential barrier shown in Fig.~1 
and discussed below, for some values of $\varepsilon_k$ (resonances) 
 the corresponding 
wave function could have   
a particular large weight inside the cluster, giving rise to deviations from 
the monotonic dependence of $\tau_A$ with energy. The potential barrier 
for Na$_{137}^-$ is 0.7 eV high. 
 
In Fig.~4 we show the distribution $P(\varepsilon)$ 
of the emitted Auger-electrons as a function
of their kinetic energies for 
Na$_{137}^-$ and Na$_{253}^-$. $P(\varepsilon)$ is calculated as 
\begin{equation}
P(\varepsilon) = \frac{A \gamma}{2 \pi} \sum_{\ell, \ell', \ell''} 
\frac{w(n \ell, n' \ell', n'' \ell'')}{(\varepsilon - \varepsilon_{k l_k})^2 
+ (\gamma/2)^2}, 
\end{equation}
 where $\varepsilon_{k l_k}(n \ell, n' \ell', n'' \ell'')$ is the energy of the
 electron emitted in the $(n \ell, n' \ell', n'' \ell'')$-Auger process, and 
 the width $\gamma$ is taking to be $0.05$ eV.  $A$ is a normalization
constant. 
 In both cluster sizes $P(\varepsilon)$ is dominated by electrons 
originated
 in  few transitions with large probability.    
 For  Na$_{137}^-$ the large peak at 1.23 eV corresponds to a 
$(1s,3p,3p)$-transition, whereas the smaller peak at 0.66 eV results from a 
$(1d,3p,3p)$-process. For Na$_{253}^-$ the Auger spectrum is dominated by 
the $(1p,2h,2h)$-process with a kinetic energy of 0.5 eV, while a smaller 
peak appears at 0.99 eV  which corresponds to the 
$(1s,4s,2h)$-transition. These two examples are
consistent with the intuitive idea that the most probable Auger processes are
those involving two electrons at the Fermi-level ($3p$ for Na$_{137}^-$, and 
 $2h$ in the case of  Na$_{253}^-$).  Results of Fig.~4 remain unchanged if 
we take into account shake-off effects.   
Note that the kinetic energies of 
the emitted electrons are, 
for all clusters studied, not smaller than 0.4 eV and not larger 
than 1.5 eV, which is roughly the difference between the bandwidth 
and the ionization potential of the clusters.  Thus, the kinetic energies 
of the Auger electrons concentrate in a narrow energy range.  Comparison with
 the experimental results by Gantef\"or et al \cite{gantefor} leads us to 
argue that the photoelectron signal which cannot be explained as coming from 
direct photoemission or thermionic effects is due to intraband Auger 
processes. 
Regarding a comparison with experimental results by Reiners and 
Haberland\cite{haberland}, one can see in Fig.~2 that the Auger life-time 
of a vacancy in the $1s$-shell, $\tau_A(1s)$, of Na$_{91}^-$ is approximately
100 ps, whereas $\tau_A(1p) \sim 10$ ps. Since the time scale 
for evaporation is probably larger than these values, our calculations 
suggest an explanation for the fact that light induced evaporation in 
Na$_{91}^-$ takes place within a photon-energy range which is smaller than 
the bandwidth.  

It is important to point out  that our results are not sensitive to the
particular form of the
functional used for the exchange and correlation term. We have found 
neither qualitative nor appreciable quantitative differences by using the 
 LDA functionals terms proposed by Wigner\cite{wigner}, 
Gunnarson-Lundqvist\cite{gl}, 
 and the parametrizations  by Perdew-Zunger\cite{perdew} and 
Vosko et al\cite{vosko} 
to the Monte-Carlo calculations of Ceperley and Alder\cite{ceperley}. 
  
We have also performed a Hartree-Fock (HF) calculation of the energy 
levels\cite{llorens}. In contrast to the results obtained using LDA, the
HF-treatment of the exchange term yields that Auger-transitions for neutral 
and positively charged clusters are energetically allowed. However, the
 number of possible transitions in Na$_N$ is much smaller than in Na$_N^-$, 
 and for Na$_N^+$ there are just a few Auger-processes. 
Thus, the HF-calculation confirms the general trends obtained using LDA.  
 
\section{Summary and Outlook}
We have calculated the intraband Auger-decay probabilities of spherical 
sodium clusters. Our results suggest that, for certain cluster sizes,  
there could be a competition between nonradiative electronic and atomic
 deexcitation channels. 
 We found that the Auger-probabilities of small metal clusters are
remarkably different from the atomic ones. 
  In view of the results presented in  this paper it remains an interesting 
 problem to study intraband Auger-probabilities  in nanostructures 
and films and to compare them with the 
case of small clusters. Also the study of spin-dependent Auger-processes in 
small magnetic clusters  appears as a possible interesting extension 
of this work. 

\section{Acknowledgements}
This work has been partially supported by the Spanish Government through the
grant PB95-0492, and by the Deutsche Forschungsgemeinschaft through the 
Schwerpunkt \lq\lq Femtosekundenspektroskopie''.

\begin{figure}[t]
\caption{
Schematic illustration of the Auger emission in a Na$_N^-$ cluster with a 
vacancy in a deep electronic shell. Note that the Kohn-Sham effective 
potential $V_{\em eff}(r)$ shows a barrier, in contrast to the case of neutral 
or positively charged clusters. The Auger probability 
for this particular process is given by the exchange  matrix element 
 $\langle k, n'' \ell'' | {\hat V} | n \ell, n' \ell' \rangle $ (see text). 
The direct Auger 
 transition is obtained by exchanging the indices $n' \ell'$ and $n \ell$.} 
\label{Figure 1}
\end{figure}

\begin{figure}[t]
\caption{
Size dependence of the Auger life-time (in picoseconds) for a hole 
in the a) first shell ($1s$), 
b) second shell ($1p$) of  spherical Na$_N^-$ clusters.}  
\label{Figure 2}
\end{figure}

\begin{figure}[b]
\caption{
Auger life-time for holes on  different shells of  
  Na$_{137}^-$. The dashed line refers to calculations taking  
into account relaxation effects for the final state.}  
\label{Figure 3}
\end{figure}

\begin{figure}[t]
\caption{
Auger electron distribution $P(\varepsilon)$ as a function of the 
electron kinetic energy  for a)  Na$_{137}^-$, and b) Na$_{253}^-$.}  
\label{Figure 4}
\end{figure}


\begin{references}
%
\item[$^*$] on leave of absence from Institut f{\"u}r Theoretische Physik, 
Freie Universit{\"a}t 
Berlin, Arnimallee 14, 14195 Berlin, Germany
\bibitem{recknagel}U. Frenzel, U. Kalmbach, D. Kreisle and E. Recknagel, 
Surf. Rev. Lett. {\bf 3}, 505 (1996).
\bibitem{bambynek} W. Bambynek, B. Crasemann, R. W. Fink, H.-U. Freund,
 H. Mark, C. D. Swift, R. E. Price and P. Venugopala Rao,  
Rev.\ Mod.\ Phys.\ {\bf 44}, 716 (1972), and references therein.
\bibitem{gantefor} G. Gantef\"or, W. Eberhardt, H. Weidele, 
D. Kreisle and E. Recknagel, Phys. Rev. Lett. {\bf 77}, 4524 (1996). 
\bibitem{haberland} Th. Reiners and H. Haberland, Phys. Rev. Lett. 
{\bf 77}, 2440 (1996).
\bibitem{walter} W. Ekardt, Phys. Rev. Lett. {\bf 52}, 1925 (1984). 
\bibitem{baladron} M. P. I\~niguez, C. Baladr\'on and J. A. Alonso,
Surf.\ Sci.\  {\bf 127}, 367 (1983). 
\bibitem{lorente} R. Monreal and N. Lorente, Phys. Rev. B {\bf 52}, 4760 
(1995). 
\bibitem{brechignac} C. Br\'echignac, Ph. Cahuzac, J. Leygnier and A. Sarfati,
Phys. Rev. Lett. {\bf 70}, 2036 (1993).
\bibitem{ekardt} W. Ekardt and Z. Penzar, Phys. Rev. B {\bf 38}, 4273 
(1988).
\bibitem{wigner} E. P. Wigner, Phys. Rev. {\bf 46}, 1002 (1934). 
\bibitem{gl} O. Gunnarson and B. I. Lundqvist, Phys. Rev. B {\bf 13}, 4274
(1976). 
\bibitem{perdew} J. P. Perdew and A. Zunger, Phys. Rev. B {\bf 23}, 5048
(1981).                                                             
\bibitem{vosko} S. H. Vosko, L. Wilk and M. Nusair, Can. J. Phys. B {\bf 58},
1200 (1980). 
\bibitem{ceperley} D. M. Ceperley and B. J.  Alder, Phys. Rev. Lett. {\bf 45},
 566 (1980). 
\bibitem{llorens} E. Lipparini, Ll. Serra and K. Takayanagi, 
Phys. Rev. B {\bf 49}, 16733 (1994). 
\end{references}
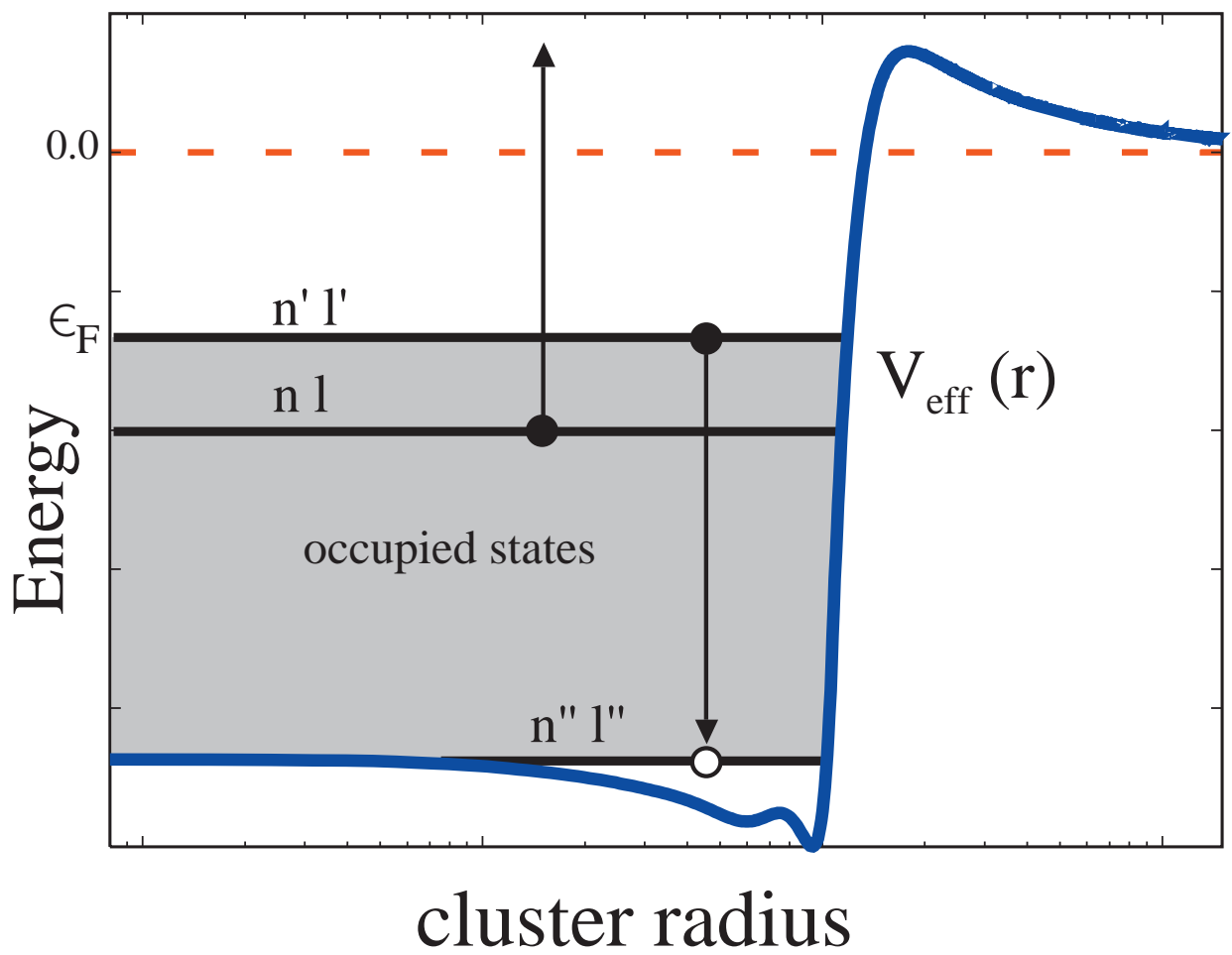
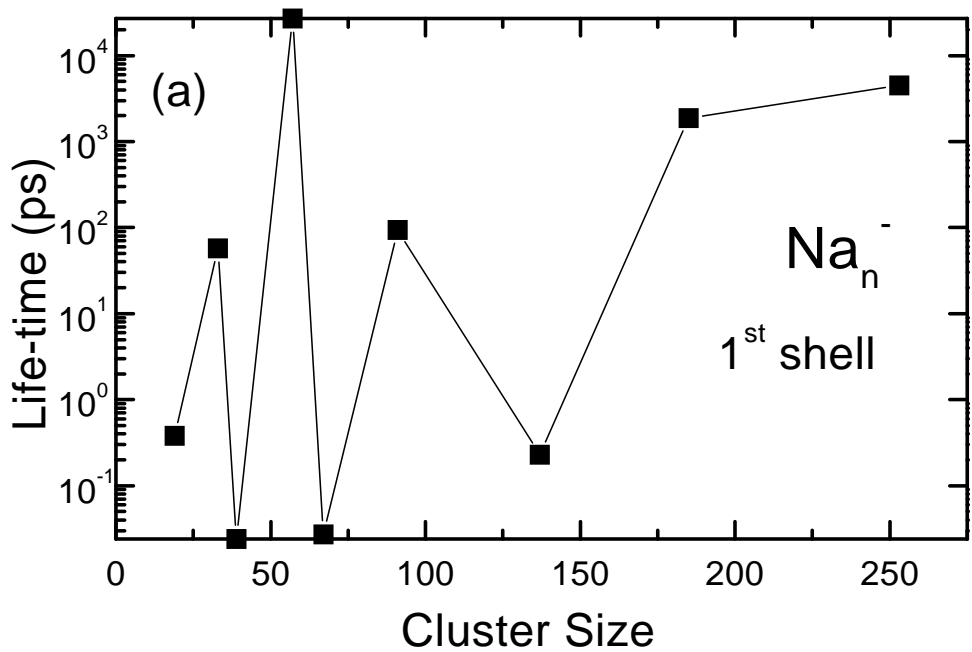
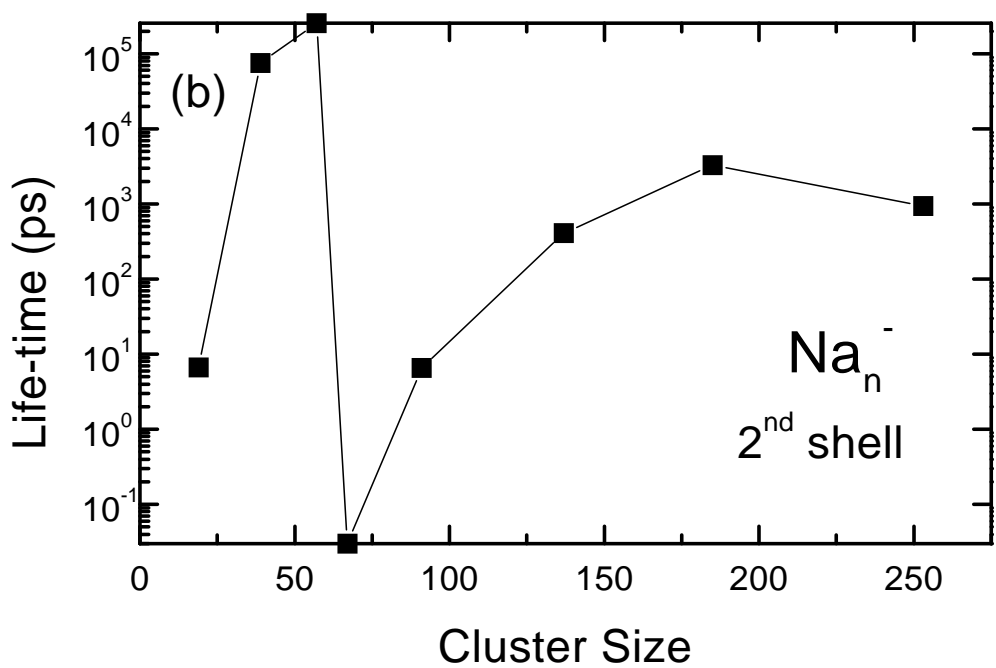
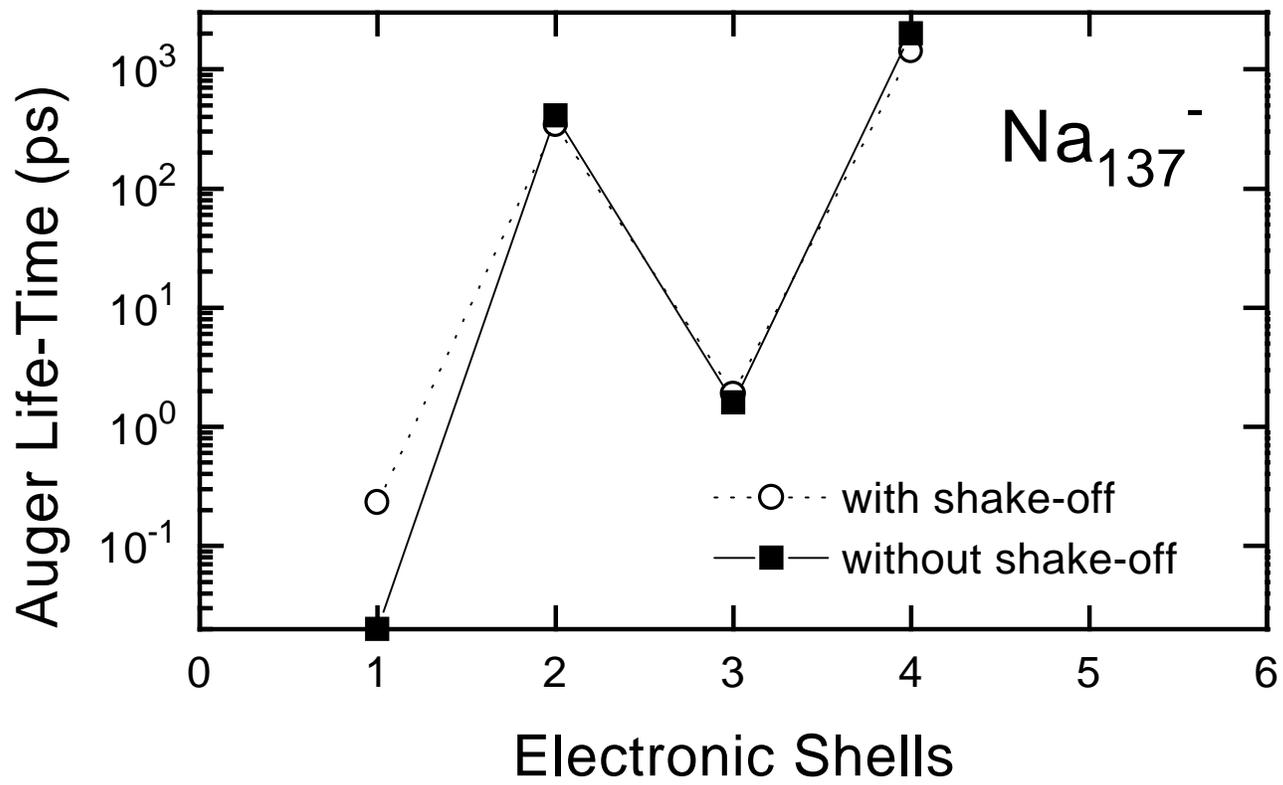
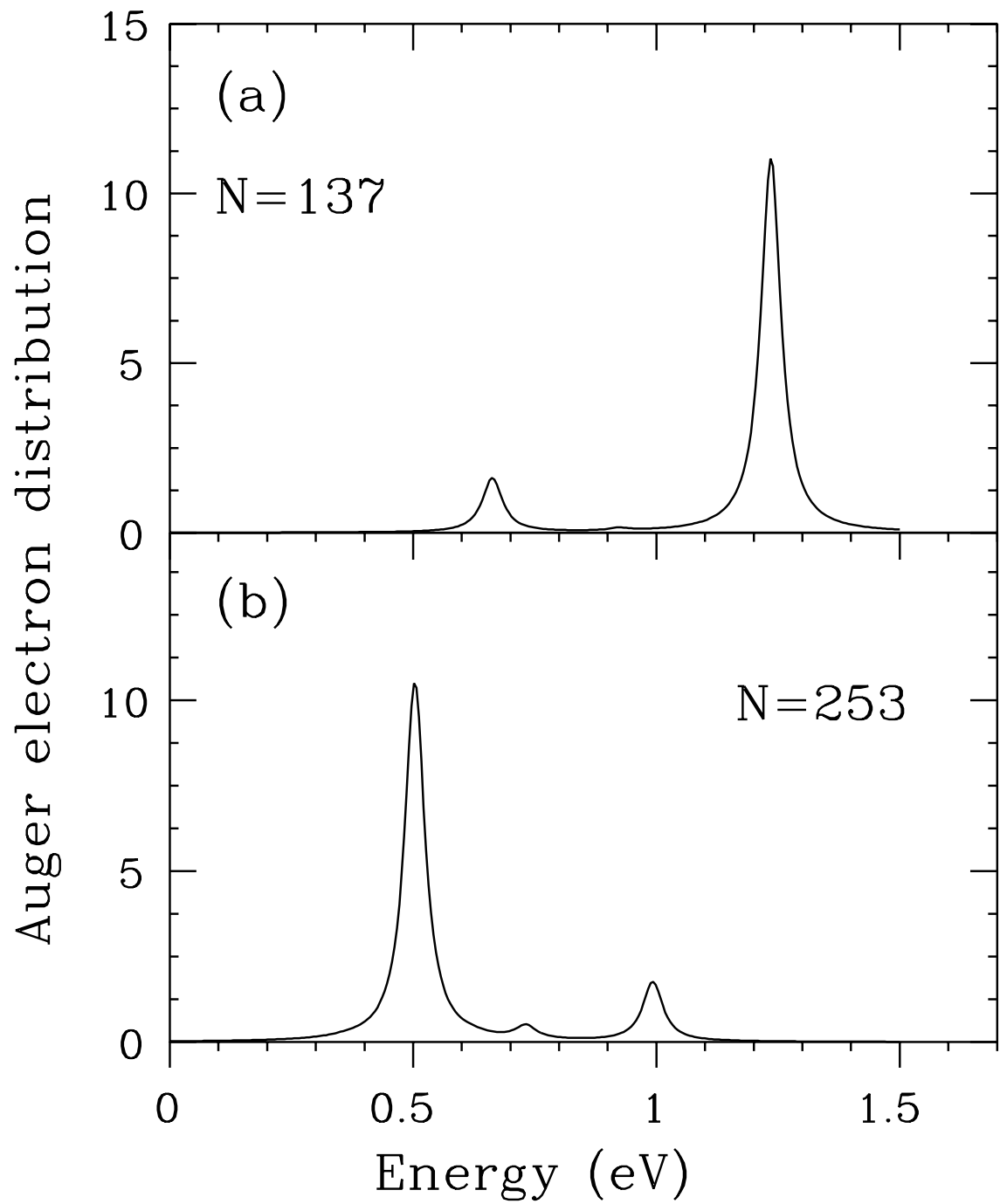
\end{document}